\documentclass[preprintnumbers,amsmath,amssymb,prc,showpacs]{revtex4} 
\usepackage[english]{babel}
\usepackage[dvips]{graphics,graphicx}
\usepackage{epsfig}
\usepackage{amssymb,amsmath,amsthm,amsfonts,amssymb}
\usepackage{dcolumn} 
\usepackage{longtable}

\begin{document}

\title{On the NUT-Born-Infeld-$\Lambda$ spacetime}
\author{Nora Bret\'on$^{}$ and C. E. Ram\'{\i}rez-Codiz$^{1}$}
\affiliation{
$^{}$ Dpto. de F\'{\i}sica, Centro de Investigaci\'on y de Estudios Avanzados 
del I. P. N., Apdo. 14-740, D.F., M\'exico.\\
$^{1}$ Fac. de F\'{\i}sica, Universidad Nacional Aut\'onoma de M\'exico
D.F., M\'exico.}

\begin{abstract}
The stationary axisymmetric spacetime coupled to nonlinear Born-Infeld electrodynamics is studied. The solution was derived by Pleba\'nski et al (1984) and it is characterized  by six free parameters:  mass, NUT charge, electric and magnetic charge, Born-Infeld parameter and cosmological constant.
The geodesic and Lorentz force equations are integrated, and a qualitative analysis of the effect of varying the parameters in the effective potential is provided. Then the light and charged particle trajectories are discussed.  The conditions that determine an extreme black hole are presented as well. 
\end{abstract}



\maketitle

\section{Introduction}

The Taub-NUT solution, first derived by Taub (1951) \cite{Taub51} and then by Newmann et al (1963) \cite{Newman63}, has been the object of intensive study because of its interesting properties.
The NUT parameter has been interpreted as the twist of an electromagnetic universe in \cite{Halilsoy2006}.
It has also been related to the gravomagnetic monopole strength and 
although addressed in many occasions, its interpretation is still in debate.
The NUT (Taub-NUT) solution is a stationary, axisymmetric one but not globally asymptotically flat because it has one semi-infinite singularity on the symmetry axis at $\theta= \pi$. Bonnor \cite{Bonnor1969} interpreted this singularity as a semi-infinite massless source endowed with a finite angular momentum. If the nondiagonal metric component is parametrized in the form $g_{t \phi}=2N \cos{\theta}$,
the interpretation involves two semi-infinite singularities located one at $\theta= 0$ and another at $\theta= \pi$; the masses of the semi-infinite sources can assume either positive or negative values \cite{Manko2005}.
Another way of solving the singularity problem was proposed by Misner \cite{Misner63} to the expense of having closed timelike geodesics. 
Interestingly, 
Lynden-Bell  and Nouri-Zonoz \cite{LBZ1998} proved that all geodesics of NUT space lie on spatial cones; this property leads to an effect of gravitational lensing: light rays are not merely bent but twisted as they pass the gravomagnetic monopole lens. The differential twisting produces a characteristic spiral shear in lensed objects, peculiar to gravomagnetic monopoles; this naturally suggests a possibility of observational detection of such an object.

Another interesting aspect related to NUT spaces has been explored in conection with
the AdS/CFT conjecture as testbeds \cite{Myers1999}. On the other hand, NUT-Reissner-Nordstrom spaces that are asymptotically dS have shown to yield counter-examples to the dS/CFT paradigm. \cite{Strominger01}. This solution has also been shown to be relevant in studies on the black hole entropy mainly in the context of Euclidean solutions or instantons \cite{Hawking1999}, \cite{Mann2006}. It has been observed as well that the entropy/area relationship does not hold in NUT spaces \cite{Mann04}, \cite{Mann2005}.

In a generalization of the Skyrme model proposed by Atiyah, Manton and Schroers \cite{Atiyah} (AMS) that aims to give a geometrical and topological interpretation to the electric charge and baryon and lepton numbers, static particles are described in terms of gravitational instantons. The electrically charged particles correspond to non-compact asymptotically locally flat (ALF) instantons. In this context the Taub-NUT instanton is a model for the electron. The geometry of the Taub-NUT skyrmion has been addressed in \cite{Dunajski}.

On the other hand, it is a well known fact that in situations involving strong electromagnetic fields, for instance fields of the order of $10^{9}$Teslas or $10^{16}$V/m, the linear superposition does not hold and nonlinear effects as the creation of electron-positron pairs is very likely to occur. 
These situations are described by Quantum Electrodynamics (QED). We also can invoke classical theories that include these nonlinear phenomena in an effective way; among these theories are the Euler-Heisenberg and the Born-Infeld (BI) nonlinear electrodynamics.  BI theory is characterized by the maximum field allowed, or BI parameter $b$, estimated of the order of $10^{20}$ Volt/m. The BI Lagrangian when expanded in electric and magnetic fields up to the order of Schwinger's, takes the form of the Euler-Heisenberg's. In particular Born-Infeld theory outstands for having several desirable properties, namely the absence of birrefringence and superluminal signals.
In the context of string theory the parameter $b$ is related to the string tension $b=1/2 \pi \alpha$ \cite{Gibbons2003}
  
For the above reasons it would be desirable to consider the NUT solution with a nonlinear electromagnetic source. This is the stationary solution to the coupled Einstein-Born-Infeld equations or, in other words, the nonlinear generalization of the NUT-Reissner-Nordstrom solution. 
In \cite{Pleban1984} the metric and the electromagnetic field of such solution were determined in terms of integrals;  the solution is equipped
with six independent parameters, including  mass, NUT charge, electric and magnetic charges, Born-Infeld parameter and cosmological constant. In this paper we present the full integration of the solution and its study in canonical coordinates that allow a simpler interpretation of the physics involved, we shall call it the NUT-Born-Infeld-$\Lambda$ (NUT-BI-$\Lambda$) solution or just NUT-Born-Infeld (NUT-BI) when  $\Lambda=0$.

There are two limiting solutions of the NUT-BI: the first one, the NUT-Reissner-Nordstrom (NUT-RN), is the linear electromagnetic limit (obtained if $b \to \infty$). The other one is the Born-Infeld (BI) solution, that is obtained if the NUT parameter is turned off.
Logically, the corresponding behaviors are different in several respects.
At the origin ($r \to 0$) the behaviour of Reissner-Nordstrom and BI is divergent to $+ \infty$; while the introduction of the NUT parameter allows to avoid this initial (spatial) singularity. 
The size of the external horizon also depends on the parameters included: it is maximum for the Schwarzschild's black hole while the introduction of electromagnetic charge makes it smaller, resulting in a denser object.
On the other hand, when the NUT parameter is introduced to the solution, has the effect of enhancing the external horizon,
whose size is between Schwarzschild's and BI's sizes.
The electric field is finite at the origin except for the RN limit.
Regarding the effective potential the BI potential barrier is higher in  comparison to the one obtained when the NUT parameter is included. These and other effects are analyzed in Section III. 
 
This paper is organized as follows:  
in Sec. II we introduce the solution and present the exact integration of the metric function;
plots of the metric function for different values of the parameters are displayed;  the corresponding electromagnetic field components are determined as well. 
In Sec. III we integrate the geodesic and the Lorentz force equations in the equatorial plane. From the expressions of the effective potentials typical trajectories of charged and uncharged test particles are presented. Light rays under the influence of nonlinear electromagnetic fields obey equations that might not coincide with null geodesics; in other words, the photon trajectories are null geodesics of an effective metric that is the background metric modified by the nonlinear electromagnetic tensor. These equations are derived and integrated and plots of the effective potentials are presented. 
In Sec. IV the conditions that determine the extreme black hole are derived, an interesting result is that the existence of the extreme black hole imposes constraints between the cosmological constant and the nonlinear Born-Infeld parameter.  Conclusions are given in the last section.

\section{The NUT-Born-Infeld solution}

In \cite{Pleban1984} all type D metrics that can be sourced by the nonlinear 
electromagnetic field of the Born-Infeld type  were derived. 
The Born-Infeld Lagrangian is given by

\begin{equation}
L_{BI}= b^2 \left\{{\sqrt{1+\frac{2F}{b^2}+\frac{G^2}{b^4}} -1}\right\}
\label{BILagr}
\end{equation}
where $F,G$ are the two electromagnetic invariants, 
$F= F_{\mu \nu}F^{\mu \nu}, \quad G=  F_{\mu \nu} \check{F}^{\mu \nu}$; 
$\check{F}^{\mu \nu}= \epsilon^{\mu \nu \alpha \beta} F_{\alpha \beta}/(2 \sqrt{-g})$ is the dual of the electromagnetic field tensor and $\epsilon^{\mu \nu \alpha \beta}$ is the Levi-Civita symbol. 
The maximum electromagnetic field of the Born-Infeld theory $b$ is the BI parameter.
In the limit that $b$ tends to infinity the linear electrodynamics limit (Maxwell) is recovered, $L=F$. Moreover, if the Born-Infeld Lagrangian is expanded for fields up to the order of Schwinger's (that are small compared with $b$) it takes the form of the Euler-Heisenberg Lagrangian. 

Among the metrics that admit a source given by Lagrangian (\ref{BILagr}) is the NUT or sometimes called Taub-NUT metric, that in the following we shall call NUT-Born-Infeld (NUT-BI) solution or NUT-Born-Infeld-$\Lambda$ (NUT-BI-$\Lambda$) when  $\Lambda \ne 0$. In the Appendix we included the metric in the coordinates $(x,y,\tau,\sigma)$ as was originally derived in \cite{Pleban1984}.
 
Changing to coordinates $(t,r, \theta, \phi)$ with $(x,y,\tau,\sigma) \mapsto (\cos{\theta},r,t,\phi)$
the NUT-BI line element can be written as

\begin{equation}
ds^2=- \psi (dt- 2N\cos{\theta} d \phi)^2+\psi^{-1}dr^2+ (N^2+r^2)[ d \theta^2+ \sin^2{\theta} d \phi^2],
\label{NUTBI2}
\end{equation} 
 
where $\psi(r)$ is given by 

\begin{equation}
\psi(r)=\frac{r^2-N^2}{r^2+N^2}-\frac{2mr}{r^2+N^2}- \frac{\Lambda}{r^2+N^2}\left({\frac{r^4}{3}+2N^2r^2-N^4}\right)
+\frac{b^2}{r^2+N^2}I(r),
\label{metric}
\end{equation}
where $m$ is the mass, $N$ is the NUT parameter, $b$ is the Born-Infeld parameter
and $\Lambda$ is the cosmological constant. The function $I(r)$ is 
related to the electromagnetic contribution and depends on the electric and magnetic charges $Q$ and $g$.

It is convenient to put the metric function $\psi$ in terms of the dimensionless variable $R=r/m$ and to identify the parameters in the following manner:
$N \to N/m$, $\Lambda \to \Lambda m^2$, $Q \to Q/m$, $b \to bm$, as well as $a^4 \to a^4/m^4=(Q^2+g^2)/b^2$; then
$\psi$ can be written as

\begin{equation}
\psi(R)=\frac{R^2-N^2}{R^2+N^2}-\frac{2R}{R^2+N^2}- \frac{\Lambda}{R^2+N^2}\left({\frac{R^4}{3}+2N^2R^2-N^4}\right)
+\frac{b^2}{(R^2+N^2)}I(R),
\label{psiR}
\end{equation}
where the integral $I(R)$ is given explicitly by
 
\begin{eqnarray}  
I(R)&=& \frac{2}{3}\left[(R^2+N^2)^2+4N^2(R^2-N^2)-(5N^2+R^2)\sqrt{(R^2+N^2)^2+a^4}\right] \nonumber\\
&& + \frac{16}{3}N^2 \sqrt{N^4+a^4}\frac{\sqrt{(R^2+N^2)^2+a^4}}{(R^2+ \sqrt{N^4+a^4})} \nonumber\\
&&+ \frac{8}{3} RN^2 (N^4+a^4)^{1/4} \left({F[\alpha,k]-2 E[\alpha,k]}\right) + \frac{2(4N^4+a^4)R}{3(N^4+a^4)^{1/4}} F[\alpha,k],
\label{Int(R)}
\end{eqnarray} 

where $a^4=(Q^2+g^2)/b^2$ and 
$F[\alpha,k]$ and $E[\alpha,k]$ are the elliptic integrals of the first and second kind, respectively. 
In Legendre's canonical form they are given by,

\begin{eqnarray}
F[\alpha,k]&=& \int_{0}^{\alpha}{\frac{d\theta}{\sqrt{1-k^2 \sin^2{\theta}}}},\nonumber\\
E[\alpha,k]&=& \int_{0}^{\alpha}{ \sqrt{1-k^2 \sin^2{\theta}} d\theta},
\end{eqnarray}
where $k$ is the modulus $0 \le k \le 1$ and $\alpha$ is the amplitude $0 \le \alpha \le \pi/2$ that in our case are,

\begin{equation}
\alpha= \arccos\left({\frac{R^2- \sqrt{N^4+a^4}}{R^2+ \sqrt{N^4+a^4}}}\right), 
\quad k^2= { \frac{\sqrt{N^4+a^4}-N^2}{2 \sqrt{N^4+a^4}}},
\label{alphak}
\end{equation}

\begin{figure}
\centering
\includegraphics[width=8cm,height=5cm]{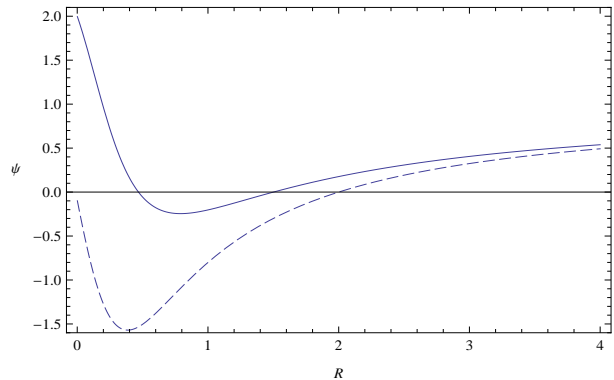}
\caption{\label{fig1}
The metric function $\psi(R)$ vs.  $R=r/m$. The plots are for different values of $Q$: The dashed curve corresponds to  $Q=0.5$ while the continuous one is with $Q=1$; as the charge augments, the exterior horizon shrinks, the same effect as in RN. The other parameters are: $b=3, \Lambda=0, N=0.5, g=0$.}
\end{figure}

\begin{figure}
\centering
\includegraphics[width=8cm,height=5cm]{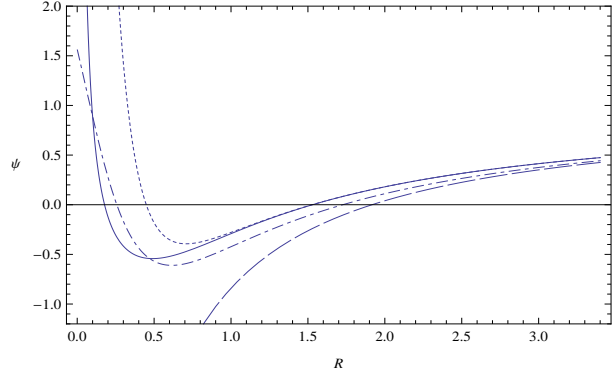}
\caption{\label{fig2} 
The metric function $\psi(R)$ is plotted as  function of the dimensionless variable $R=r/m$.
The continuous plot corresponds to the RN-like behaviour  ($b=1.3,N=0,Q=0.85$); the dashed one is the Schwarszchild-like   ($b=1.3,N=0,Q=0.4$); dot-dashed is for the NUT-BI solution ($b=4.5,N=0.5,Q=0.85$); the dotted plot is the BI case ($b=4.5,N=0,Q=0.85$). Notice the stretching on the size of the exterior horizon when the NUT parameter is introduced, compared to the BI horizon size. In all the plots $\Lambda=0, g=0$.}
\end{figure}

We note that to plot Elliptic functions in MATHEMATICA, you should take the second argument squared.
The function $I(R)$ depends on the parameters $b,Q,g$ and $N$, that clearly shows the NUT charge interaction with the electromagnetic field. Some plots of the metric function (\ref{psiR}) are shown in Figs \ref{fig1} and \ref{fig2}. In Fig. \ref{fig1} it is shown the effect of changing the charge $Q$ ($g=0$); as $Q$ grows the horizon shrinks, the black hole becoming a more compact object, an effect similar to the one in Reissner-Nordstrom. In Fig. \ref{fig2}, plots of $\psi$ are shown to compare the NUT-BI behavior with the Schwarzschild, RN and BI cases.
 

\subsection{Electromagnetic fields}

Regarding the electromagnetic field, from the electromagnetic two-form  Eq. (\ref{emtwo-form}) we derive the nonvanishing components of the electromagnetic tensor: 

\begin{equation}
F_{tr}=E, \quad F_{\theta \phi}=(N^2+r^2)B, \quad F_{r \phi}=2N \cos{\theta} E.  
\end{equation}
 
The explicit expressions for the electric field $E$ and the magnetic flux density $B$ are

\begin{equation}
E_{NUT-BI}=\frac{Q \cos{\Phi}-g \sin{\Phi}}{\sqrt{a^4+(r^2+N^2)^2}},
\label{Electric} 
\end{equation}

\begin{equation}
B_{NUT-BI}= \frac{Q \sin{\Phi}+g \cos{\Phi}}{(r^2+N^2)},
\label{Magnetic}
\end{equation} 

where the function $\Phi$ is given by

\begin{equation} 
\Phi= -2N \int_{r}^{\infty}{\frac{ds}{ \sqrt{(s^2+N^2)^2 + a^4}}}=
- \frac{N}{(a^4+N^4)^{1/4}} F\left[\alpha, k \right],
\end{equation}

where $\alpha$ and $k$ are given in Eqs. (\ref{alphak}) and $a^4=(Q^2+g^2)/b^2$.  
$B$ and $E$ are related to the magnetic field $H$ and the electric displacement $D$ through the structural equations,

\begin{equation} 
D= \exp{[\nu(r)]}E, \quad H= \exp{[-\nu(r)]}B
\end{equation}
with $\exp{[\nu(r)]}$ given by
\begin{equation}
\exp[2 \nu]= 1+ \frac{Q^2+g^2}{b^2(r^2+N^2)^2},
\label{nu}
\end{equation}

The explicit expressions of  $D$ and $H$ are

\begin{eqnarray} 
D&=& \frac{Q \cos (\Phi) - g \sin (\Phi) }{N^2+r^2}, \nonumber\\
H&=& \frac{Q \sin (\Phi) + g \cos (\Phi) }{\sqrt{a^4+(N^2+r^2)^2}},
\end{eqnarray}

From Eq. (\ref{Electric}) 
we can get to the limiting cases electric fields,

\begin{eqnarray}
E_{NUT-RN}&=&\frac{Q (r^2-N^2)+2gNr}{(r^2+N^2)^2}, \nonumber\\
E_{BI}&=&  \frac{\sqrt{Q^2+g^2}}{\sqrt{a^4+r^4}},\nonumber\\
E_{RN}&=&  \frac{\sqrt{Q^2+g^2}}{r^2}.
\end{eqnarray}

\begin{figure}
\centering
\includegraphics[width=8cm,height=5cm]{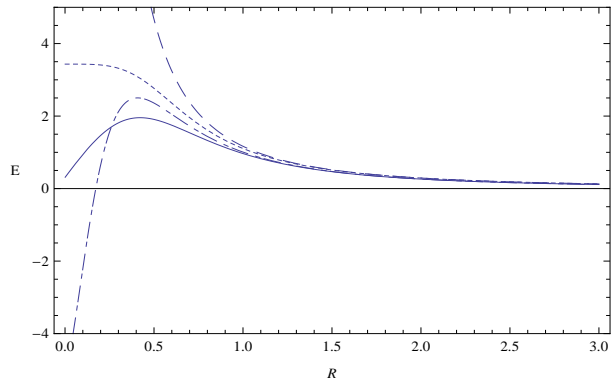}
\caption{\label{fig4} 
The electric field $E$ is shown for different values of the parameters; the continuous plot is the NUT-BI case ($b=2.5,N=0.4,Q=0.85,g=0.8$); the divergent field is the Reissner-Nordstrom (RN) case (dashed $Q=0.85, g=0.8$); dot-dashed is for the NUT-RN limit ($N=0.4,Q=0.85, g=0.8$) and the dotted plot is the BI case ($b=2.5,N=0,Q=0.85,g=0.8$)}
\end{figure}

Plots of the electric field and its limiting cases are shown in Fig. \ref{fig4}. For cases NUT-BI, NUT-RN and BI
the electric field is finite at the origin; RN diverges at $r=0$. Introducing the NUT parameter, $E$ becomes less intense.

Regarding the magnetic field, from the general expression Eq. (\ref{Magnetic})
we can get the limit cases,

\begin{eqnarray}
B_{NUT-RN}&=&\frac{ g(r^2-N^2)-2QNr}{(r^2+N^2)^2}, \nonumber\\
B_{BI}&=&  \frac{g}{{r^2}}, \nonumber\\
B_{RN}&=&  \frac{g}{r^2}, 
\end{eqnarray}
\subsection{Cosmological constant}

\begin{figure}
\centering
\includegraphics[width=8cm,height=5cm]{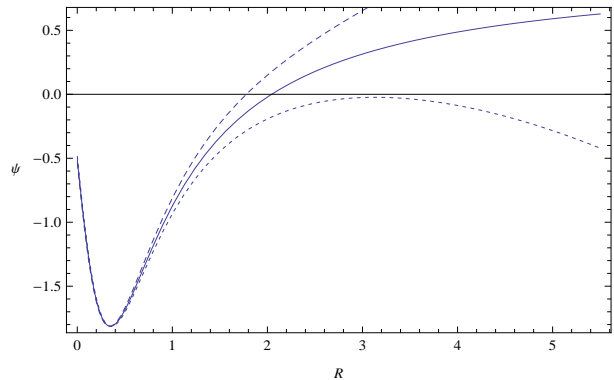}
\caption{\label{fig3}
The effect of including the cosmological constant $\Lambda$ with different signs in
the metric function $\psi(R)$. It is plotted as function of the dimensionless variable $R=r/m$. The dotted curve is for $\Lambda=0.1$ (de Sitter), the dashed one corresponds to  $\Lambda=-0.1$ (Anti--de Sitter) while the continuous one is for  $\Lambda=0$. The values of the set of parameters are: $b=1.3,Q=0.4, N=0.5, g=0$. }
\end{figure}

The inclusion of the  cosmological constant has interest in black hole thermodynamics; Hawking and Page \cite{HawkingPage1983} showed the existence of a certain phase transition in the phase space of a Schwarzschild-Anti-de Sitter black hole; and recently the Anti-de Sitter parameter has been interpreted as a thermodynamic pressure, that jointly with its conjugate quantity is considered a more complete description of the thermodynamic phase space \cite{Mann2012} in which the Smarr relation arises in a natural way. The effect of including $\Lambda$ in the metric function is shown in Fig \ref{fig3}. If the sign of $\Lambda$ is positive, it corresponds to the de Sitter case; and for a critical $\Lambda$, the black hole could disappear, if the metric function become negative for all values of the radial coordinate. For a negative  $\Lambda$ (Anti-de Sitter), the horizon is smaller than the one with $\Lambda=0$.

\subsection{On the singularity and asymptotics}

From Eqs. (\ref{curvature}) we are able to explore if the spacetime is singular. The only nonvanishing  Weyl scalar is $C^{(3)}$ ($\Psi_2$ in the Penrose notation \cite{Ernst}) and any invariant must be constructed from it. 
We can easily prove, just by doing the indicated derivatives, that the Weyl scalar does not diverge; neither diverge the scalar curvature $R$. Then this is a locally regular solution.
 
Regarding the asymptotic of the NUT-BI solution with $\Lambda=0$, it corresponds to the so called asymptotically locally flat (ALF) spaces, whose study, associated to the NUT solution, has been addressed in several papers, the interest being mainly related to instantons and black hole thermodynamics, see for instance \cite{Mann2005}.
\subsection{Limit cases}
 
In the limit case with $N=0$
we recover the Born-Infeld (BI) solution \cite{breton1}, characterized by the parameters $m, b, Q, g$; given by the line element Eq. (\ref{metric}) making $N=0$  and the function $I(r)$ given by

\begin{equation}
I_{BI}(r)= \frac{2}{3a^4r}(r^4-r^2 \sqrt{r^4+a^4}) + \frac{2}{3a} F\left[\arccos{\left({\frac{r^2-a^2}{r^2+a^2}}\right)}, \frac{1}{\sqrt{2}}\right].
\end{equation}
  
This is the case without cosmological constant, $\Lambda=0$; it can be included in the solution
just by keeping the third term with $\Lambda$ in the metric (\ref{metric}).   
From this solution we can obtain Reissner-Nordstrom (RN) by making $b \to \infty$; and then we can obtain Schwarzschild's if $Q=0, g=0$.

On the other hand, the linear limit of NUT-BI is obtained if $b \to \infty$ in which case we recover the NUT- Reissner- Nordstrom (NUT-RN) solution with $\Lambda \ne 0$, given by the line element \cite{Mann2006b}

\begin{equation}
\psi(r)=\frac{r^2-N^2}{r^2+N^2}-\frac{2mr}{r^2+N^2}- \frac{\Lambda}{r^2+N^2}\left({\frac{r^4}{3}+2N^2r^2-N^4}\right)
+\frac{Q^2+g^2}{r^2+N^2},
\label{metric2}
\end{equation}
with the electromagnetic potential given by

\begin{eqnarray}
A_{t}=- \frac{Qr}{r^2+N^2}+ \frac{g({r^2-N^2})}{2N(r^2+N^2)} \nonumber\\
A_{\phi}= \frac{2NQr-g({r^2-N^2})}{r^2+N^2} \cos {\theta},
\end{eqnarray}

This solution was first derived in \cite{Brill1964} as a homogeneous cosmology with electromagnetic field.
The supersymmetric properties of
this solution has been studied in \cite{Alberca2000}.
From this solution making $N=0$ we recover the RN solution and then if $Q=0$ we get Schwarzschild's.
Making directly in metric (\ref{metric2}) $Q=0, g=0$ we arrive to NUT solution characterized by the mass parameter and NUT charge, $(m,N)$. The typical behavior of $ \psi$ is shown in Fig. \ref{fig2}.
To locate the NUT-BI-$\Lambda$ solution in the context of exact solutions of Einstein equations with electromagnetic source we include the Table \ref{T:1} (see also \cite{Kramer} Ch. 21).

{\renewcommand{\tabcolsep}{2.mm}
{\renewcommand{\arraystretch}{1.}
\begin{table}[htbp]
\begin{minipage}{0.8\textwidth}
\caption{\label{T:1} 
In the table are shown the characteristic independent parameters of the Einstein-Maxwell solutions as well as the Born-Infeld nonlinear generalizations derived in \cite{Pleban1984}.
The parameters marked by a cross ($\times$) are different from zero in the corresponding solutions. They are: 
the mass $m$, NUT parameter   $N$, electric charge $Q$, magnetic charge $g$,  cosmological constant $\Lambda$  and BI denotes the Born-Infeld parameter $b$. The 
Pleba\'nski (1975), Demia\'nski-Newman (1966) and  Carter (1968) solutions
possess in addition a rotation parameter. }
\centering
\resizebox*{0.75\textwidth}{!}{
\begin{tabular}{cccccc|c}
\multicolumn{6}{c}{} &  \multicolumn{1}{c}{} \\
\hline
\hline
$m$  & $N$&$Q$&$g$ & $\Lambda$ & BI& References \\ \hline \hline
 $\times$ &$\times$ &$\times$& $\times$& $\times$& $\times$& NUT-BI-$\Lambda$;  Pleba\'nski (1984) \cite{Pleban1984} \\ 
 $\times$ & & $\times$&$\times$ &$\times$ &$\times$ & BI-$\Lambda$; Pleba\'nski (1984)  \cite{Pleban1984} \\ 
 & &$\times$ &$\times$ &$\times$ &$\times$ &   BR-BI-$\Lambda$;Pleba\'nski (1984) \cite{Pleban1984}\\ 
  $\times$ &$\times$ &$\times$ &$\times$ &$\times$ & & Pleba\'nski (1975) \cite{Pleban75}\\  
 $\times$ &$\times$ &$\times$ &$\times$ & & & Demia\'nski-Newman (1966) \cite{DemNewman66} \\ 
$\times$ &$\times$ & $\times$& &$\times$ & & Carter (1968) \cite{Carter68}\\ 
 $\times$ & $\times$&$\times$ & & & &   NUT-RN; Brill (1969) \cite{Brill1964}\\
 $\times$ & &$\times$ & & & &   RN; Reissner-Nordstrom \\ 
 & &$\times$ &$\times$ & & &   BR; Bertotti-Robinson (1959)\\\hline \hline
\end{tabular}}
\end{minipage}
\end{table}}}
  
\section{Geodesics and charged particle trajectories}

In this section we determine the geodesics and the effect of varying the parameters in the effective potential.
To integrate the geodesics we shall use two of the existing motion conserved quantities. 
The geodesic equation is given by

\begin{equation}
\frac{d^2x^{\alpha}}{d \tau^2} + \Gamma^{\alpha}_{\beta \delta} \frac{dx^{\beta}}{d \tau} \frac{dx^{\delta}}{d \tau}=0,
\label{geodesic}
\end{equation}
where $\tau$ is the affine parameter that generates the geodesics. 
When integrating it is helpful to consider the invariant obtained from the line element (\ref{NUTBI2}),

\begin{equation}
\delta= - \psi (\dot{t}-2N \cos{\theta}\dot{\phi})^2 +\psi^{-1}\dot{r}^2+ (N^2+r^2)(\sin^2{\theta}\dot{\phi}^2+\dot{\theta}^2),
\end{equation}
where dot denotes derivative w.r.t. the affine parameter $\tau$ and
$\delta=-1$ for massive particles and $\delta=0$ for massless particles. 
Moreover, the spacetime possesses at least two Killing vectors associated to the cyclic coordinates $t$ and $\phi$, that give rise to two constants of motion that we identify as  the energy and angular momentum per unit mass of the test particle, $\varepsilon$ and $L$, respectively,

\begin{eqnarray}
\varepsilon&=& \psi (-\dot{t}+ 2N \cos{\theta} \dot{\phi}), \nonumber\\
L&=& \psi [2N \cos{\theta} \dot{t} - 4N^2 \cos^2{\theta} \dot{\phi}] + (N^2+r^2) \sin^2 {\theta} \dot{\phi}.
\end{eqnarray}

Substituting in the geodesic equation the values of $\Gamma^{\alpha}_{\beta \delta}$ as well as $\varepsilon$ and $L$ we get for the $r$ coordinate $x^{\alpha}=r$,

\begin{equation}
\ddot{r}- \delta \frac{\psi'}{2}+   \left({\frac{L+2N \cos{\theta} \varepsilon}{\sin{\theta}}}\right)^2 \frac{d}{dr}\left({\frac{\psi}{2(N^2+r^2)}}\right)-\left({\frac{\psi r}{\sin^2{\theta}}-\frac{\psi'}{2}(N^2+r^2)}\right) \dot{\theta}^2=0.
\end{equation}

This equation can be integrated in the equatorial plane, $\theta=\pi/2, \dot{\theta}=0$, using that it can be written as

\begin{equation}
\frac{d}{d \tau}\left({\dot{r}^2- \delta \psi}+ \frac{L^2 \psi}{(N^2+r^2)}\right)=0,
\end{equation}

that integrating leads to

\begin{equation}
\dot{r}^2- \delta \psi+ \frac{L^2 \psi}{(N^2+r^2)} + {\rm const}=0,
\label{rdot}
\end{equation}

Identifying the effective potential by comparison with $\dot{r}^2 + V_{\rm eff}^2=$const,

\begin{equation}
V_{\rm eff}^2= \psi \left( \frac{L^2}{(N^2+r^2)} - \delta \right).
\end{equation}

\begin{figure}
\centering
\includegraphics[width=8cm,height=5cm]{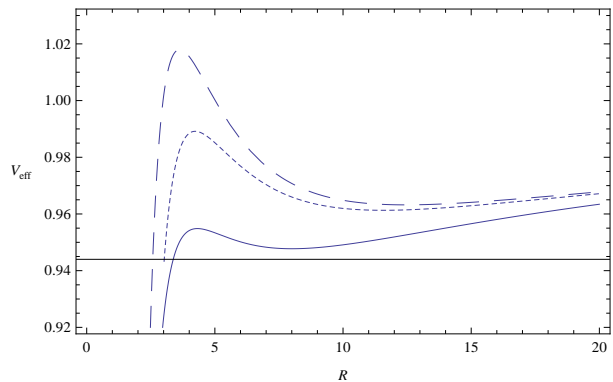}
\caption{\label{fig5} 
The effective potential $V_{\rm eff}$ plotted as a function of $R=r/m$, for different values of the angular momentum $L$ of uncharged test particle ($\delta=-1$). The continuous curve is for $L=3.5$ while the dashed one is for $L=4$.
The values of the other parameters are: $b=3,Q=0.5, \Lambda=0, N=0$. The dotted curve is the effective potential for $L=4$ with NUT parameter ($N=0.5$), the inclusion of $N$ lowers the potential barrier.}
\end{figure}

\begin{figure}
\centering
\includegraphics[width=8cm,height=5cm]{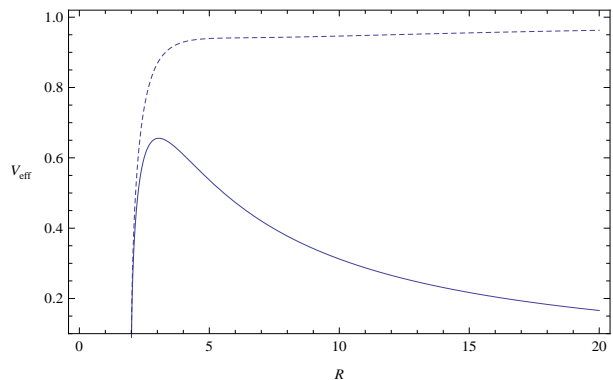}
\caption{\label{fig6}
In the plot it is shown the effective potential $V_{\rm eff}^2= \psi( {L^2}/{(N^2+r^2)} - \delta )$ for uncharged particles. Continuous plot is for massless particles  $\delta=0$; while dot-dashed is the one felt by massive particles $\delta=-1$. In both plots $b=3,N=0.5,Q=0.5,L=3.5$.}
\end{figure}

In Figs. \ref{fig5}, \ref{fig6} and \ref{fig7} are shown some of the effects of varying the parameters of the black hole $ b, Q, N, \Lambda$ and the angular momentum of the test particle, $L$; we shall fix $g=0$.

In Fig. \ref{fig5} the shapes of the effective potential for uncharged test particles are displayed. The effect of diminishing their angular momentum $L$ is to lowering the potential barrier. For the same value of the angular momentum, the effect of introducing the NUT parameter $N$ is to lowering the potential barrier, meaning that a greater NUT mass makes the black hole more attractive. Moreover, as the NUT parameter grows, the potential barrier is less intense and the size of the exterior horizon is enhanced. From the plots it is also evident that there are bounded orbits as well as circular orbits, both stable (minimum of $V_{\rm eff}$) and unstable (maximum of $V_{\rm eff}$).

In Fig. \ref{fig6} the different effective potentials for massive and massless test particles with the same angular momentum $L$ are shown. Massless particles are less likely to be captured by the black hole.

\begin{figure}
\centering
\includegraphics[width=8cm,height=5cm]{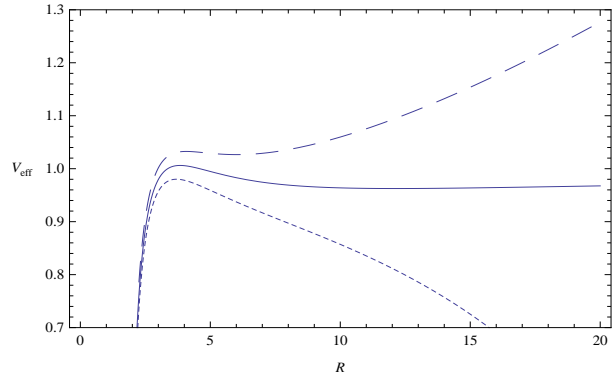}
\caption{\label{fig8}
The effect in the effective potential of including the cosmological constant $\Lambda$ with different signs.  
$V_{\rm eff}$ is plotted as function of the dimensionless variable $R=r/m$. The dotted curve is for $\Lambda=0.005$ (de Sitter), the dashed one corresponds to  $\Lambda=-0.005$ (Anti-de Sitter) while the continuous one is for  $\Lambda=0$. The values of the set of parameters are: $b=3,Q=0.5, N=0.3, L=4$. Compare with the effect of varying $\Lambda$ on the metric function $\psi(R)$ shown in Fig. \ref{fig3}  }
\end{figure}

The effect in
the effective potential of including the cosmological constant $\Lambda$ with different signs is shown in Fig \ref{fig8}. A nonzero cosmological constant changes completely the black hole nature. Equilibrium positions of test particles associated to the minimum of $V_{\rm eff}$ dissapear. This behavior can be compared with the corresponding to the metric function shown in Fig. \ref{fig3}. 



\subsection{Charged test particles trajectories}

To describe the trajectories of charged test particles we must consider the Lorentz force equation,

\begin{equation}
\frac{d^2x^{\alpha}}{d \tau^2} + \Gamma^{\alpha}_{\beta \delta} \frac{dx^{\beta}}{d \tau} 
\frac{dx^{\delta}}{d \tau}=
\left({\epsilon F^{\alpha}_{\cdot \nu }+i \gamma (L_{,F}\check{F}^{\alpha}_{\cdot \nu}+ L_{,G}{F}^{\alpha}_{ \cdot \nu })}\right) 
\frac{dx^{\nu}}{d \tau},
\label{LorentzEq}
\end{equation}

where $\epsilon$ and $\gamma$ are the electric and magnetic charge of the test particle. 
Note that the l.h.s. of the previous equation when equated to zero is  the geodesic equation  Eq. (\ref{geodesic}) that we have just integrated. 
The right hand side of the Lorentz equation  can be integrated at the equatorial plane as well. For $x^{\alpha}=r$, the r.h.s. of Eq. (\ref{LorentzEq}) reduces to,

\begin{eqnarray}
\left({\epsilon F^{r}_{\cdot \nu }+i \gamma (L_{,F}\check{F}^{r}_{\cdot \nu }+ L_{,G}{F}^{r}_{\cdot \nu })}\right) \dot{x}^{\nu}&=&
(- \psi \dot{t}) \epsilon E+i \gamma (- \psi \dot{t})iH = \nonumber\\
&& = \frac{\epsilon \varepsilon }{2N} \frac{d}{dr}\left({Q \sin{\Phi}+g \cos{\Phi}}\right) -
\frac{\gamma \varepsilon}{2N} \frac{d}{dr}\left({-g \sin{\Phi}+Q \cos{\Phi}}\right),
\end{eqnarray}
   
In this case the conserved energy has a term coming from the interaction with the electromagnetic field, $\varepsilon= - \partial_{t}^{a}(m U_{a}+ \epsilon A_{a})$.
Now defining the constants $\Delta_g=g \epsilon - Q \gamma$ and  $\Delta_q=Q \epsilon + g \gamma$, multiplying by $\dot{r}$ to integrate it as a function of the affine parameter $\tau$, and considering Eq. (\ref{rdot}),  we arrive to 

\begin{equation}
\dot{r}^2= \delta \psi - \frac{L^2 \psi}{(N^2+r^2)} + \frac{\varepsilon}{N} \left[{\Delta_q  \sin{\Phi}+ \Delta_g \cos{\Phi}}\right] + {\rm const},
\end{equation}

identifying the effective potential as

\begin{eqnarray} 
V_{\rm eff}^2&=& \psi \left({\frac{L^2}{(N^2+r^2)}-\delta}\right) +   V_{\rm em}+ {\rm const} \nonumber\\
&=& \psi \left({\frac{L^2}{(N^2+r^2)}- \delta}\right) - \frac{\varepsilon}{N} \left[{\Delta_q  \sin{\Phi}+ \Delta_g \cos{\Phi}}\right] + {\rm const}
\end{eqnarray}

\begin{figure}
\centering
\includegraphics[width=8cm,height=5cm]{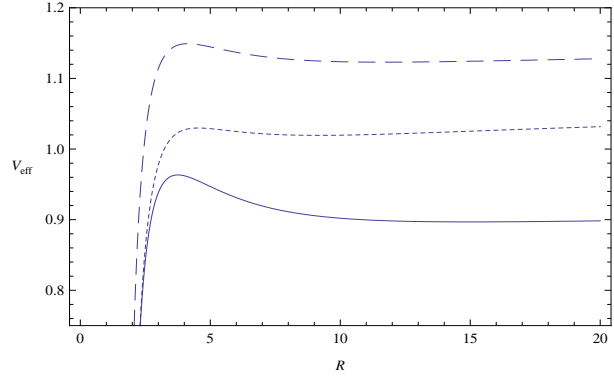}
\caption{\label{fig9}
The effective potential for charged particles $V_{\rm eff}^2= \psi( {L^2}/{(N^2+r^2)}+1 )-V_{\rm em}$ is plotted. The continuous plot corresponds to a test particle with electric and magnetic charges $\epsilon=-0.15, \gamma=0.1$,
while the dashed one is the effective potential with the opposite charge $\epsilon=0.15$ and the same magnetic charge; finally the dotted plot corresponds to a magnetic charge $\gamma=-0.1$ and $\epsilon =-0.15$. For all plots $b=1.5,Q=0.6,g=1,N=0.5, \varepsilon=0.8, \Lambda=0$ and $L=4$. }
\end{figure}

The constant should be determined from the condition that at $r \to \infty, \quad V_{\rm eff} \to 0$. 
In Fig. \ref{fig9} it is shown the effect on the effective potential of varying the parameter $N$;
the greater is the NUT parameter the lower the potential barrier, that is a similar effect for uncharged particles, see Fig. \ref{fig5}.
 

\subsection{Light trajectories in the effective NLED metric}
 
In Maxwell electrodynamics, light trajectories are those of the null geodesics and the characteristic surfaces $S$ along which the discontinuities of the electromagnetic field propagate, are  given  by

\begin{equation}
g^{\mu \nu}S_{,\mu}S_{,\nu}=0.
\label{Nullgeo}
\end{equation} 
   
However if strong electromagnetic fields are involved, 
nonlinear effects will arise and the characteristic surfaces obey the null geodesic equation modified by the electromagnetic tensor \cite{nordita},

\begin{equation}
\left({g^{\mu \nu}+ \frac{4 \pi}{b^2}T^{\mu \nu}}\right) = \gamma^{\mu \nu}S_{,\mu}S_{,\nu}=0,
\end{equation} 
these are the null geodesics of an effective metric, $\gamma_{\mu \nu}$.
Clearly in the Maxwell limit, $b \mapsto \infty$, Eq. (\ref{Nullgeo}) is recovered.
  
For the NUT-BI spacetime the nonvanishing components of the energy-momentum  tensor are

\begin{eqnarray}
8 \pi T^{\phi}_{\phi}&=& 8 \pi T^{\theta}_{\theta}=\Lambda +2b^2 (e^{- \nu}-1)= \Lambda +2b^2 \left({\frac{r^2+N^2}{\sqrt{(r^2+N^2)^2 + a^4}}-1}\right), \nonumber\\
8 \pi T^{t}_{t}&=&8 \pi T^{r}_{r}= \Lambda +2b^2 (e^{\nu}-1)= \Lambda +2b^2 \left({\frac{\sqrt{(r^2+N^2)^2 + a^4}}{r^2+N^2}-1}\right), \nonumber\\
T^{t}_{\phi}&=&2N \cos{\theta}[T^{\phi}_{\phi}-T^{t}_{t}],
\end{eqnarray}
where $\nu$ is given in Eq. (\ref{nu}). Substituting these expressions into the line element of the effective metric, $\gamma^{\mu \nu}$, for $\delta=0$, we obtain,

\begin{equation}
\left({\frac{\Lambda}{2b^2} + \frac{\sqrt{(r^2+N^2)^2 + a^4}}{r^2+N^2}}\right)[ -\psi (\dot{t} - 2 N \cos{\theta}\dot{\phi})^2 + \psi^{-1} \dot{r}^2]+\left({\frac{\Lambda}{2b^2} + \frac{r^2+N^2}{\sqrt{(r^2+N^2)^2 + a^4}}}\right)(N^2+r^2)( \sin^2{\theta}\dot{\phi}^2 + \dot{\theta}^2)=0,
\label{null_eff_geod}
\end{equation}

From the previous equation some comments are in order. 
In the case that $a=0$, the factor in brackets can be factorized from all the expression becoming a conformal factor. The nonlinear field distort the light trajectories as long as $a \ne 0$, i.e. when $Q \ne 0$.
Moreover, there is a twisting effect in two situations: if $\dot{\phi} \ne 0$, there is a twisting  
associated to the angular momentum of the test particle. If  $\dot{\theta} \ne 0$, i.e. in any trajectory out of the equatorial plane a similar effect occurs. In the case of radial equatorial trajectories, i.e.  both $\dot{\phi} = 0$ and $\dot{\theta} = 0$, this dragging effect will not appear.
  
On the equatorial plane, ${\theta} = \pi /2$ Eq. (\ref{null_eff_geod}) can be reduced to

\begin{equation}
\dot{r}^2+ \frac{ \chi \psi L^2}{(N^2+r^2)}= \varepsilon^2,
\label{light_traj}
\end{equation}

where $\chi$ is the term that arises from the nonlinear interaction and determines how much the light trajectories deviate from the null geodesics (for which $\chi=1$); $\chi$ is given by
\begin{equation} 
\chi= \frac{\Lambda+2b^2e^{- \nu}}{\Lambda+2b^2e^{\nu}} =\frac{\Lambda \sqrt{a^4+(N^2+r^2)^2}+ 2b^2 (N^2+r^2)}{\Lambda \sqrt{a^4+(N^2+r^2)^2}+ 2b^2 (N^2+r^2)+2(Q^2+g^2)/(N^2+r^2)},
\end{equation}
 
\begin{figure}
\centering
\includegraphics[width=8cm,height=5cm]{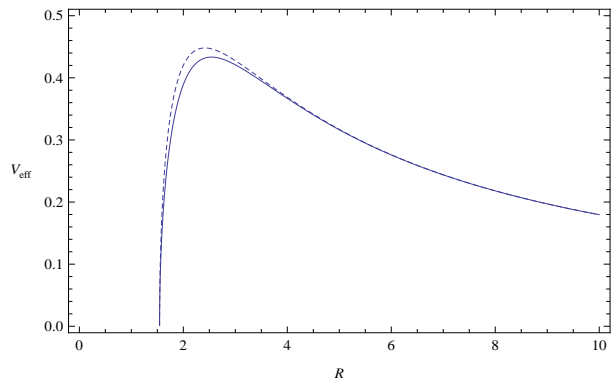}
\caption{\label{fig10}
Effective potentials for massless particles and photons.
The dotted plot is the effective potential felt by massless particles,
$V_{\rm eff}^2= \psi {L^2}/{(N^2+r^2)}$; while continuous plot is the effective potential felt by photons $V_{\rm eff}^2= \chi \psi {L^2}/{(N^2+r^2)}$; for both plots $b=0.75,N=0.1,Q=0.85,L=2, \Lambda=0$}.
\end{figure}

From Eq. (\ref{light_traj}) we can identify the effective potential in the second term of the l.h.s. 
The term $ \chi$ is absent in the equation for the null geodesics of the metric $g_{\mu \nu}$.
Provided $\Lambda \ge 0$, it is easy to show that $ 0 < \chi \le 1$; therefore the effective potential for photons is lower than the one for massless particles, effect that one can think of as a screening of the charge, in such a manner that photons that would not be trapped by the black hole, if nonlinear electromagnetic effects are taken into account, they  fall down the hole. The effect when $\Lambda=0$ is shown in Fig. \ref{fig10}.
The case when $\Lambda <0$ (deSitter) corresponds to an odd behaviour: there are light rays that do never cross the horizon, since it might be that $\chi \to 0$ and then $V_{\rm eff}=0$, effect that might be interpreted as if light rays do not feel the presence of the black hole.

In the case that $a=0$ there is no distortion of the null geodesics and light trajectories coincide with the former. The linear behaviour is recovered if $b \mapsto \infty$ ($a$ in its turn goes to zero and $\chi=1$) and then the null geodesics coincide with those of the NUT-RN spacetime. 

The orbits on the plane $(r, \phi)$, $r( \phi)$ can also be determined, using that
$\dot{r}/\dot{\phi}= dr / d \phi$, 

\begin{equation}
d \phi= dr \left[{\frac{\varepsilon^2}{L^2}(N^2+r^2)^2- \chi \psi (N^2+r^2)}\right]^{-1/2}
\end{equation}

while the deflection angle  of the light trajectories in the cone of the NUT spacetime is given by
$\alpha=2 \Delta \phi - \pi$.


\section{The NUT-BI-$\Lambda$ Extreme Black Hole}
   
For the extreme Reissner-Nordstrom (RN) black hole the two horizons, $r_{\pm}= M \pm \sqrt{M^2-Q^2}$, coalesce into one i.e. it is characterized by one parameter, $M=Q$.
It is well known that such a system is of interest because it is related to the BPS states that are those states in string theory that preserve symmetries; its thermodynamics also has arisen controversy, since it is a stable system and emits no Hawking radiation, as its temperature should be zero. 
In this section
the conditions that are needed in order that the two horizons of the NUT-BI-$\Lambda$  black hole
coalesce into one are determined. Without loss of generality $g=0$ will be assumed. 
\begin{figure}
\centering
\includegraphics[width=8cm,height=5cm]{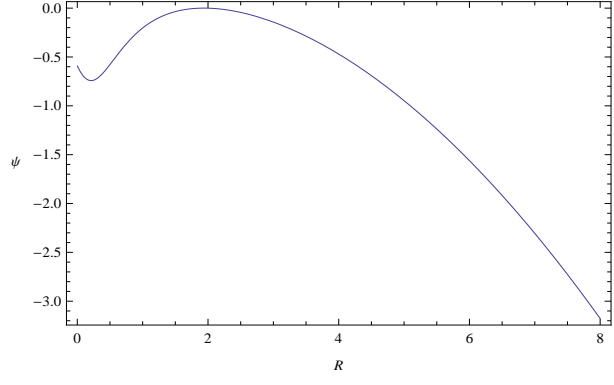}
\caption{\label{fig11}
de Sitter case: the metric function is plotted for a case where $\psi(r)=0$ and $\psi(r)'=0$. 
Despite the conditions for a unique horizon are fulfilled, one can hardly call it a \textit{black hole}. The values of the parameters are $Q=1.5, b=30.117, \Lambda=0.18, N=0.615$.}
\end{figure}

The extreme black hole is characterized by one horizon, $r_h$; at that horizon $r_h$ both the metric function $\psi=0$ and its first derivative $\psi'=0$ vanish.
We can use that the derivative of $I(r)$ from Eq. (\ref{Int(R)}) is 

\begin{equation}
I'(r)= - \frac{2(r^2+N^2)}{r^2[r^2+N^2+ \sqrt{(r^2+N^2)^2+a^4}]}=\frac{2(r^2+N^2)}{r^2a^4}[r^2+N^2- \sqrt{(r^2+N^2)^2+a^4}].
\end{equation}  

Using the condition $\psi (r_h)=0$ we can write  $I(r)$ at the horizon $r_h$ as, 

\begin{equation}
I(r_h)= \frac{1}{Q^2 r_h}\left\{{N^2 -r_h^2 +2mr_h + \Lambda \left({\frac{r_h^4}{3}+2N^2r_h^2-N^4}\right)}\right\}.
\end{equation}

Then the first derivative of $\psi(r)$ can be written as

\begin{eqnarray}
\psi'(r)&=& \frac{1}{3(r^2+N^2)^2}\left[ 6m(r^2-N^2)+12N^2r +\Lambda(-2r^5-4N^2r^3-18N^4r) \right] \nonumber\\ &&+Q^2 \left[\frac{(N^2-r^2)}{(r^2+N^2)^2}I(r)+ \frac{rI'(r)}{(r^2+N^2)}\right],
\end{eqnarray}
that when evaluated at $r_h$ gives

\begin{equation}
\psi'(r)= \frac{1}{r_h}- \frac{\Lambda}{r_h}(r_h^2+N^2)+\frac{2b^2}{r_h}
\left[r_h^2+N^2- \sqrt{(r_h^2+N^2)^2+a^4}\right].
\end{equation}

From the condition $\psi'(r)=0$ a quadratic equation for $(N^2+r^2)$ can be obtained.
The cases  $\Lambda > 0$, $\Lambda = 0$ and $\Lambda < 0$ shall be addressed separately.

\subsection{de Sitter Case ($\Lambda > 0$)}

From the conditions above  the value of the extreme horizon in terms of $\Lambda >0, Q, b$ and $N$ is determined by

\begin{equation}
r_h^2+N^2= \frac{2b^2- \Lambda  \pm  \sqrt{b^4 -4b^4 \Lambda Q^2+ b^2 \Lambda^2 Q^2}}{4 \Lambda b^2-\Lambda^2}.
\label{extrrh}
\end{equation}

In order to have only one horizon radious, the square root in Eq. (\ref{extrrh}) must be zero, this condition gives a constraint for  $Q^2$,

\begin{equation}
Q^2 = \frac{b^2}{\Lambda (4b^2- \Lambda)}.
\label{cond1}
\end{equation}
 
Note that the denominator in Eq. (\ref{cond1}) should be positive, then $4b^2 > \Lambda$; moreover, once the square root in Eq. (\ref{extrrh}) is zero, its numerator must be positive, what leads to a more stringent condition, $2b^2 > \Lambda$ that includes the former. Regarding the NUT parameter, an additional condition must be satisfied,

\begin{equation}
r_h^2= \frac{2b^2- \Lambda}{4 \Lambda b^2-\Lambda^2} -N^2>0,
\label{extrrh2}
\end{equation}
that can be fulfilled with a NUT parameter small enough. Therefore we have shown that the extreme case cannot exist in an arbitrary de-Sitter space, there are three constraints in order to fulfill that $\psi=0$ and $\psi'=0$ and these conditions do not fix uniquely a set of parameters but only define a region in the parameter space. Even if those conditions are satisfied, one can hardly call the resulting system a black hole; the case is shown in Fig \ref{fig11} for some particular values of the paarmeters. The negative cosmological constant produces a negative metric function that cannot be interpreted as a stationary spacetime, but
as some kind of \textit{cosmology}. Remind that when $\psi (r) < 0$ the $r$ coordinate  becomes timelike while the time coordinate should be interpreted as spacelike.

\subsection{$\Lambda=0$ Case}

In the case that $\Lambda=0$ an analogous derivation leads to

\begin{equation}
r_h^2+N^2= Q^2- \frac{1}{4b^2},
\label{BIext}
\end{equation}
that leads to a lower bound for the charge $Q$,

\begin{equation} 
Q^2 \ge N^2+ \frac{1}{4b^2},
\label{lam_zero}
\end{equation} 
the Eq. (\ref{BIext})  is the BI extreme black hole condition identifying $r^2 \mapsto (r^2+N^2)$ \cite{Chemissany2008}.
In the limit $b \to \infty$, corresponding to the NUT-Reissner-Nordstrom limit, the condition is $Q^2 \ge N^2$. Besides,
Therefore, in this case there is a constraint on the parameters given by Eq.(\ref{lam_zero}), but such constraint does not define uniquely the parameter values for an extreme black hole. Other conditions fixing the rest of parameters are needed; for instance, to fix another parameter it could be used $\psi'(r_h)=0$ or $\psi(r_h)=0$, but degeneracy persists.
  
\subsection{AdS Case ($\Lambda <0$)}

It shall be considered $\Lambda <0$ from the very beginning in the metric function $\psi(r)$, so we rename the cosmological term
as $l^2=-3/ \Lambda $, and the metric function reads as

\begin{equation}
\psi(r)=\frac{r^2-N^2}{r^2+N^2}-\frac{2mr}{r^2+N^2}+ \frac{({r^4}+6N^2r^2-3N^4)}{l^2(r^2+N^2)}
+\frac{b^2}{r^2+N^2}I(r).
\label{metricAdS}
\end{equation}

From the conditions $\psi=0$ and $\psi'=0$ and
following an analogous procedure we arrive to the quadratic equation for $(N^2+r^2)$ 

\begin{equation}
\left({\frac{9}{l^4}+ \frac{12b^2}{l^2}}\right)(N^2+r^2)^2+\left({\frac{6}{l^2}+ 4b^2}\right)(N^2+r^2)+(1-4Q^2b^2)=0,
\end{equation}
whose roots are

\begin{equation}
r^2+N^2= x_{\pm}=\frac{-l^2(3+2l^2b^2) \pm 2 \sqrt{9b^2l^4Q^2+b^4l^6(l^2+12Q^2)}}{9+12b^2l^2}.
\label{rootsAdS}
\end{equation}
 
Being physically meaningless the negative root $x_{-}$ is discarded. The other one, $x_{+}$, must be positive, and this requirement leads to the constraint that the numerator in Eq.(\ref{rootsAdS}) must be positive, that in turn leads to the constraint that $4b^2Q^2>1$ and moreover, that $x_{+}-N^2>0$, that amounts to

\begin{equation}
l^4(3+4b^2l^2)(4Q^2b^2-1)>N^2(3+4Q^2)[2l^2(3+2b^2l^2)+3N^2(3+4Q^2)]>0,
\end{equation}
that can be fulfilled with an apropriate (small) value of $N$.
Therefore in this AdS case there are two conditions to be fulfilled by the parameters of the solution, that again do not define a unique set of parameters defining an extreme black hole. 


It would be very interesting to explore the thermodynamic aspects of this extreme black hole, mainly those related to the breakdown of the usual relation between entropy and  area of the horizon \cite{Mann04}, \cite{Mann2006}. 
The expression for the surface gravity $\kappa$ can be calculated as
$\kappa= \frac{1}{2} g_{tt}'(r_h)$,

\begin{equation}
\kappa=\frac{1}{2r_h}- \frac{\Lambda}{2r_h}(r^2+N^2)+\frac{b^2}{r_h}\left[r_h^2+N^2- \sqrt{(r_h^2+N^2)^2+a^4}\right].
\end{equation}
It is believed that an extreme black hole has zero temperature, $T= \kappa / 2 \pi$.
The values of $r_h$ that make $\kappa=0$ can be determined by solving 
$\kappa(r_h)=0$, that is a quadratic equation in $(r^2+N^2)$; for $\Lambda \ne 0$,

\begin{equation} 
(r_h^2+N^2)= \frac{\Lambda- 2b^2 \pm {2b^2} \sqrt{1-4 \Lambda Q^2+ \Lambda^2 a^4}}{ \Lambda^2 - 4 \Lambda{b^2}}.
\end{equation}
   
Expression that coincides with the condition for the extreme black hole Eq. (\ref{extrrh}).
In the case that $\Lambda = 0$, the condition amounts to $(r^2+N^2)= Q^2-{1}/{4b^2} $, that is the BI case if we identify $r^2 \mapsto (r^2+N^2)$. The extreme case deserves a thorough study that however is beyond the scope of this article.
  
\section{CONCLUSIONS}

We fully integrated the solution presented by Pleba\'nski et al \cite{Pleban1984}, for a stationary axisymmetric spacetime sourced by the nonlinear electromagnetism of Born-Infeld. The solution is characterized by the parameters of NUT charge $N$, mass $m$, electric and magnetic charge ($Q,g$), cosmological constant $\Lambda$ and Born-Infeld parameter $b$. We obtained the limiting cases: NUT-Reissner-Nordstrom (NUT-RN) and  Born-Infeld (BI), and made a comparison of the corresponding electric and magnetic fields.
  
The geodesic and the Lorentz force equations were integrated on the equatorial plane ($\theta= \pi/2$) and plots on the effect of the parameter variation in the effective potential were presented. We conjecture that
it should be possible to integrate the test particle trajectories also outside of the equatorial plane $\theta \ne \pi/2$, on the basis that the NUT-RN spacetime do possess
four Killing vectors and one Killing tensor (see for instance \cite{NUT-RN}), that we believe the nonlinear generalization preserves; therefore, the existence of more constants of motion should allow the full integration of the geodesics.
The effective nonlinear electromagnetic metric whose null geodesics are the trajectories of light rays was determined as well.
  
Conditions for the existence of the extreme black hole were derived in the form of ranges in the space of parameters; interestingly some constraints between the charges, the BI parameter and the cosmological constant were obtained, that however do not determine a unique set of parameters for an extreme black hole, being six parameters of
the solution degeneracy exists in defining the extreme solution.  
Among other results, we learned that the extreme black hole cannot exist in an arbitrary de Sitter space.

Regarding how the lensing effect changes in NUT spaces when the nonlinear field is introduced we do not expect that BI field could make a difference, since lensing is a large distance effect. Moreover,  it would be difficult to distinguish between NUT-BI and NUT-RN in the trajectories for charged particles at long distances, but differences should be noticed near the horizon.
 
It would be interesting to classify the allowed orbits in the effective potentials in a meticulous way
(see for instance \cite{Claus2010}), as well as
the behavior of the effective potential with varying cosmological constant and the trajectories that arise are worth of a thorough study. The interesting parameters included in the NUT-BI-$\Lambda$ solution makes the thermodynamics of the system be also a promising subject that however is beyond the scope of this work.
 


\section{Appendix}
In \cite{Pleban1984} all type D solutions of the coupled Einstein-Born-Infeld equations were derived
using the null tetrad formalism. To this end, the two directions defined by the electromagnetic field were aligned along the two Deveber-Penrose vectors. In this work the null tetrad formalism is not used, but rather  a coordinate system that renders more physical insight.
 
For completeness we include the solution as was presented in the null tetrad formalism \cite{Ernst}. The line element is given in terms of the null tetrad $e^{a}, a=1,2,3,4,$

\begin{equation}
g= 2 e^{1}\otimes e^{2} + 2 e^{3}\otimes e^{4}, \quad  2 e^{1} = \bar{e}^{2}, \quad  2 e^{3}= e^{4}
\end{equation}

and the two-form of the nonlinear electromagnetic field is given by

\begin{equation}
\omega= \frac{1}{2}(F_{ab}+\check{P}_{ab})  e^{a}\wedge e^{b}.
\end{equation} 

The closure condition of the two-form $d \omega =0$ is equivalent to the Maxwell-Faraday equations

\begin{equation}
\check{F}^{ab}_{;b}=0, \quad {P}^{ab}_{;b}=0.
\end{equation}

The system of Equations for nonlinear electrodynamics is closed by the Einstein equations

\begin{equation}
R_{ab}- \frac{1}{2}g_{ab}R= 8 \pi T_{ab}+ \Lambda g_{ab},
\end{equation}

The object of our study, the nonlinear electromagnetic generalization of the NUT-Carter metric, or the Born-Infeld generalization of the NUT solution in coordinates $(x,y, \tau, \sigma)$ is given by

\begin{equation}
ds^2=(l^2+y^2)\left({Pd \sigma^2 +\frac{dx^2}{P}}\right)+ \frac{l^2+y^2}{\tilde{Q}} dy^2 -\frac{\tilde{Q}}{l^2+y^2}(d \tau -2lx d \sigma)^2,
\label{NUTBI1}
\end{equation} 
where the metric functions are given by

\begin{eqnarray}
P(x)&=& \alpha+ \beta x- \epsilon x^2,\nonumber\\
\tilde{Q}(y)&=& \epsilon (y^2-l^2)-{2my}- \lambda (y^4/3+2l^2y^2-l^4)  +(e^2+g^2)y I(y)\nonumber\\
I(y)&=& \int_{y}^{\infty}{\frac{ds}{s^2} \frac{2}{1+ \sqrt{1+(e^2+g^2)/(b^2(s^2+l^2)^2)}}},
\end{eqnarray}

The parameters are: $\alpha$, $\beta$ and $\epsilon$; the latter can be parametrized as $0,1,-1$;
$m$ is the gravitational mass, $l$ is the NUT parameter, $e^2$ and $g^2$ are the
electric and magnetic charges, respectively; $\lambda$ is the cosmological constant
and $b$ is the Born-Infeld parameter.

The electromagnetic two-form is given by:

\begin{equation}
\omega= \frac{1}{2il}(e+ig)d\left\{ { \exp(i \Phi)(d \tau-2lx d\sigma)}\right\},
\label{emtwo-form}
\end{equation}

where
\begin{equation}
\Phi =  \varphi_0-2l \int_{y}^{\infty}{ds \left[{(s^2+l^2)^2 + \frac{(e^2+g^2)^2}{b^2}}\right]^{-1/2}},
\end{equation}
 
The complement of the given solution are the curvature quantities, that in the notation of Pleba\'nski \cite{Ernst}
are given by

\begin{eqnarray}
C^{(3)}&=& - \frac{2}{(y+il)^3} \left[{m+il\left({\epsilon - \frac{4}{3}\lambda l^2}\right)}\right] + \nonumber\\
&& + \frac{1}{6}\frac{(y+il)^3}{(y^2+l^2)} \partial_{y} \partial_{y} \left({\frac{(e^2+g^2)y}{(y+il)^3}}
\int_{y}^{\infty}{\frac{ds}{s^2} \frac{2}{1+ \sqrt{1+(e^2+g^2)/(b^2(s^2+N^2)^2)}}}\right), \nonumber\\
C_{12}&=& - \frac{e^2+g^2}{(y^2+l^2)^2} \left({1+ \frac{a^4}{(y^2+l^2)^2}}\right)^{-1/2}, \nonumber\\
R&=& -4 \lambda -4b^2\left[{\left({1+ \frac{a^4}{(y^2+l^2)^2}}\right)^{-1/2}+\left({1+ \frac{a^4}{(y^2+l^2)^2}}\right)^{1/2} -2}\right].  
\label{curvature}
\end{eqnarray}

To study the solution 
We have changed coordinates $(x,y,\tau,\sigma) \mapsto (\cos{\theta},r,t,\phi)$; fixing the parameters
$\alpha=1$, $\beta=0$, $\epsilon=1$, $\varphi_0=0$, and renaming some constants ($l \to N$, $e \to Q$, $\lambda \to \Lambda$ ) we arrive to Eq. (\ref{NUTBI2}).  

\begin{acknowledgments}
C.E.R.C. acknowledges a CONACyT (Mexico) fellowship.
N. B.  acknowledges partial support of CONACyT (Mexico), project 166581.
\end{acknowledgments}

\end{document}